\documentclass[%
 aip,
 amsmath,amssymb,
 reprint,%
floatfix,
]{revtex4-1}

\usepackage{amsfonts}
\usepackage{amsmath}
\usepackage{amssymb}
\usepackage{graphicx}
\usepackage{mathrsfs}
\usepackage{color}
\usepackage{bbold}
\usepackage{CJK}
\usepackage{times}
\usepackage[colorlinks, citecolor=red]{hyperref}

\usepackage{graphicx}
\usepackage{dcolumn}
\usepackage{bm}

\usepackage[utf8]{inputenc}
\usepackage[T1]{fontenc}
\usepackage{mathptmx}
\usepackage{color}

\begin{document}

\title[]{Gate-control of the current-flux relation of a Josephson  quantum interferometer based on proximitized metallic nanojuntions}

\author{Giorgio De Simoni}
\email{giorgio.desimoni@sns.it}
\affiliation{NEST, Istituto Nanoscienze-CNR and Scuola Normale Superiore, I-56127 Pisa, Italy}
\author{Sebastiano Battisti}
\affiliation{NEST, Istituto Nanoscienze-CNR and Scuola Normale Superiore, I-56127 Pisa, Italy}
\affiliation{Department of Physics ”E. Fermi”, Università di Pisa, Largo Pontecorvo 3, I-56127 Pisa, Italy}
\author{Nadia Ligato}
\affiliation{NEST, Istituto Nanoscienze-CNR and Scuola Normale Superiore, I-56127 Pisa, Italy}
\affiliation{TeCIP Institute, Scuola Superiore Sant’Anna, 56124 Pisa, Italy}
\author{Maria Teresa Mercaldo}
\affiliation{Dipartimento di Fisica “E. R. Caianiello”, Università di Salerno, Fisciano, Salerno IT-84084, Italy}
\author{Mario Cuoco}
\affiliation{SPIN-CNR, Fisciano, Salerno IT-84084, Italy}
\affiliation{Dipartimento di Fisica “E. R. Caianiello”, Università di Salerno, Fisciano, Salerno IT-84084, Italy}
\author{Francesco Giazotto}
\email{francesco.giazotto@sns.it}
\affiliation{NEST, Istituto Nanoscienze-CNR and Scuola Normale Superiore, I-56127 Pisa, Italy}

\preprint{AIP/123-QED}

\begin{abstract}
We demonstrate an Al superconducting quantum interference device in which the Josephson junctions are implemented through gate-controlled proximitized Cu mesoscopic weak-links. The latter behave analogously to genuine superconducting metals in terms of the response to electrostatic gating, and provide a good performance in terms of current-modulation visibility. 
We show that, through the application of a static gate voltage, we are able to modify the interferometer current-flux relation in a fashion which seems  compatible with the introduction of $\pi$-channels within the gated weak-link. Our results suggest that the microscopic mechanism at the origin of the suppression of the switching current in the interferometer is apparently phase coherent, resulting in an overall damping of the superconducting phase rigidity. 
We finally tackle the performance of the interferometer in terms of responsivity to magnetic flux variations in the dissipative regime, and discuss the practical relevance of gated proximity-based all-metallic  SQUIDs for magnetometry at the nanoscale.

\end{abstract}

\maketitle

\begin{figure}[ht]
  \includegraphics[width=\columnwidth]{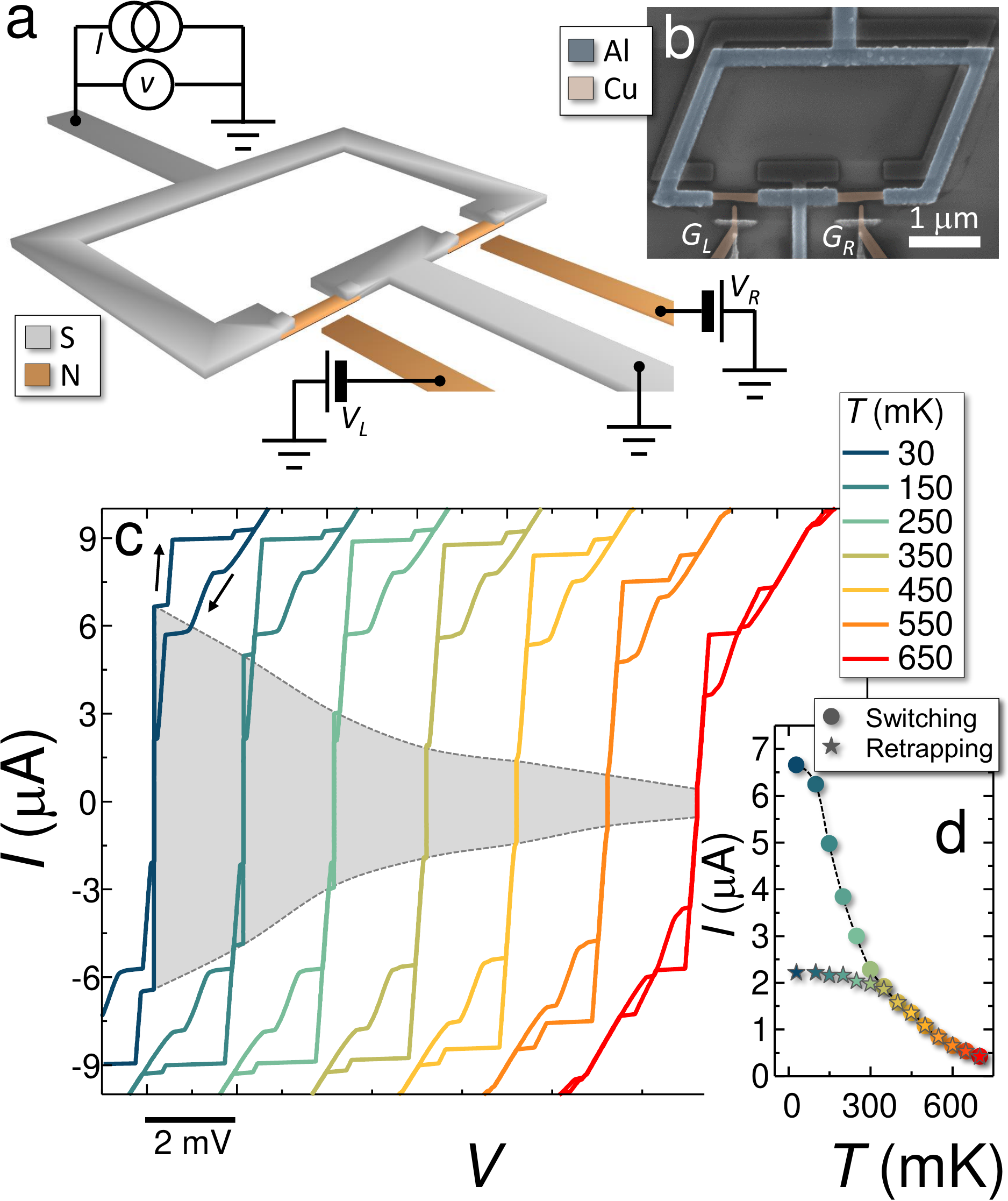}
    \caption{\textbf{Proximity-based gated all-metallic DC  SQUID} \textbf{a}: Scheme of a DC superconducting quantum interference device (SQUID) based on superconductor/normal metal /superconductor (SNS) gated proximity Josephson junctions. The 4-wire electrical setup is also shown. The left (L) and right (R) gates are polarized with voltage $V_L$ and $V_R$, respectively. \textbf{b}: False-color scanning electron micrograph of a representative gated SNS SQUID. The Al interferometer ring is coloured in blue. The Cu Josephson weak-links and the left ($G_L$) and right ($G_R$) are colored in orange. \textbf{c}: Current ($I$) \textit{vs.} voltage ($V$) forward and backward characteristics of a representative SNS SQUID at selected temperatures between 30 mK and 650 mK and at magnetic flux $\phi=0.135 \phi_0$. Curves are horizontally offset for clarity. The $I-V$ region corresponding to the presence of a Josephson current is coloured in gray. \textbf{d}: Switching (dots) and retrapping (stars) current of the same device of panel c \textit{versus} temperature $T$. The difference between switching and retrapping current stems from  heat generated in the junctions when approaching the superconducting state from the dissipative regime.}
  \label{fig:fig1}
\end{figure}
All-metallic gated superconducting transistors (GSTs) are a class of mesoscopic quantum devices, entirely realized with Bardeen-Cooper-Schrieffer (BCS) metals, in which the critical supercurrent ($I_C$) can be largely regulated via electrostatic gating \cite{DeSimoni2018,Paolucci2018,Paolucci2019b,Paolucci2019,DeSimoni2020,Puglia2021,Rocci2021}. Differently from proximitized semiconductors and low charge-density superconductors \cite{Nishino1989a,Mannhart1993b,Mannhart1993a,Okamoto1992a,Fiory1990a,Akazaki1995,Doh2005a}  where the critical current is controlled via conventional field-effect-driven charge-density modulation, in GSTs the $I_C$ suppression is obtained, regardless of the sign of the gate voltage, without the carrier concentration being affected \cite{Virtanen2019a}. The underlying physical mechanism has not been clearly identified yet, and a few hypotheses have been guessed to explain a plethora of experimental results, which cannot be comprehended in the bare framework of the BCS theory \cite{Tinkham2004}.

Recently, a high-energy electron injection due to cold electron field emission from the gate, has been claimed \cite{Ritter2021,Alegria2021,Golokolenov2020,Ritter2021a} to have a major role in $I_C$ suppression. This picture does not rely on novel physics. Yet, it does not seem compatible with some of the observed phenomenology such as the absence of a sum rule between currents originating from different gates \cite{Paolucci2019},
the response of in-vacuum suspended gated superconducting nanowires \cite{Rocci2020}, 
and the non-thermal character of the switching current probability distributions of GSTs\cite{Puglia2020,Puglia2021a}.
A possible alternative \cite{Amoretti2020,Solinas2021,Schwinger1951} explanation relies on the analogy between the creation of an electron-positron couple from the vacuum by a constant electric field in quantum electrodynamics (i.e., the so-called Sauter-Schwinger effect),  and the creation of an excited condensate in a BCS superconductor. 
As another choice, the involvement of a voltage-driven orbital polarization at the surface of the superconductor has been proposed \cite{Mercaldo2020,Mercaldo2020a,Bours2020a} to be responsible for an unconventional phase reconstruction of the superconducting order parameter, leading to weakening and destruction of  superconductivity. While high-energy electron injection due to field emission is likely to be strongly detrimental for preserving phase coherence in the superconductor, the two latter models are supposed to preserve it up to a large extent, and both predict the occurrence  of a rotation of $\pi$ in the macroscopic superconducting phase of the region affected by the gate voltage.

The information about the phase behavior of a superconductor subjected to the action of external stimuli can be experimentally accessed through a DC superconducting quantum interference device (SQUID)\cite{Clarke2004}: a superconducting ring interrupted by two Josephson weak links in parallel. A magnetic field threading the loop controls the current \textit{vs.} voltage ($IV$) characteristics of the SQUID via magnetic flux quantization \cite{Doll1961,Deaver1961} and the DC Josephson effect\cite{Josephson1962}, thus resulting in a modulation of the amplitude of the critical supercurrent. The impact of the electrostatic gating on the superconducting phase of a BCS superconductor was investigated so far only in monolithic Ti interferometers based on gated Dayem bridges \cite{Paolucci2019a}. 
Such systems allowed to retrieve a footprint of the action of the gating on the switching current ($I_S$) \textit {vs.} flux ($\phi$) relation  of the SQUID. 
Nonetheless, due to the large value of the SQUID inductance, the $I_S(\phi)$ of these interferometers exhibited poor modulation visibility, with a significant deviation from the ideal sinusoidal behavior \cite{Clarke2004}. This limited the access to a detailed information on the dependence of the current \textit{vs.} phase relation  of gated metallic Josephson weak links on the applied voltage.

Here we tackle this relevant question by demonstrating a SQUID in which the Josephson junctions are implemented through gate-controlled   Cu weak-links. 
These can   carry a dissipationless phase-dependent supercurrent thanks to the proximity effect \cite{Pannetier2000} induced by the superconducting Al forming the interferometer ring \cite{Ronzani2013,Ronzani2014,Angers2008}. Gated superconducting/normal-metal/superconducting (SNS) proximitized Josephson weak links, based on Al/Cu/Al junctions, were recently demonstrated to behave analogously to genuine superconducting metals in terms of the response to electrostatic gating \cite{DeSimoni2019}. Furthermore, this kind of weak-links possess typically a Josephson inductance significantly larger than superconducting Dayem bridges, securing a good performance in terms of current-modulation visibility \cite{Ronzani2013,Ronzani2014,Angers2008}. For the above reasons, we selected such system as the suitable candidate to explore the impact of gating on the CPR of a metallic weak link. 
Specifically, we show that the application of a constant gate voltage results into a strongly modified SQUID current-flux relation that might be compatible with the occurrence of a frustration of the superconducting phase due to activation of $\pi-$ domains within the weak-link. In addition, we discuss the performance of gated proximity-based all-metallic SQUIDs in terms of responsivity to magnetic flux variations in the dissipative regime.

\section{Effect of gate voltage on the SQUID current-flux relation}
Our  gate-controllable superconducting interferometers (SNS SQUID) consist of a 100-nm-thick Al superconducting loop interrupted by two Al/Cu/Al  planar gated  junctions. The loop of the SQUID spans a surface of  about 7.5 $\mu$m$^2$. Aluminum shows a strong proximization capability over copper,  thanks to the good quality of the interfaces formed between these two metals \cite{DeSimoni2019}. 
The  Cu  normal-metal  wire was  120 nm  wide, 630 nm long,  and  20  nm  thick. The weak links operate in the diffusive regime and within the long-junction limit, holding when the Thouless energy of the junction $E_{Th}=\frac{\hbar D}{L^2}\simeq 13$ $\mu$eV$ \ll \Delta_{Al} \simeq 180$ $\mu$eV, where $D\simeq0.008$ m$^2$/s is the Cu diffusion coefficient \cite{DeSimoni2019},  $L$ the weak-link length, and $\Delta_{Al}$ the superconducting gap of the Al banks. 
Moreover, two  80-nm-wide Cu gate electrodes, labelled  G$_L$ and G$_R$, were separated from the normal-metal wire by a distance of about 60 nm and 45 nm, respectively (in the representative device whose data are discussed in the following). Further  details  of the fabrication  process  are reported in  the \textit{Methods} section. A 3-dimensional representation of a typical SNS SQUID comprising the scheme of the 4-wire electrical setup is displayed in  Fig. \ref{fig:fig1}a,  whereas  a  false  color  scanning electron micrograph  of a representative device is shown in Fig.  \ref{fig:fig1}b.
\begin{figure}[ht]
  \includegraphics[width=\columnwidth]{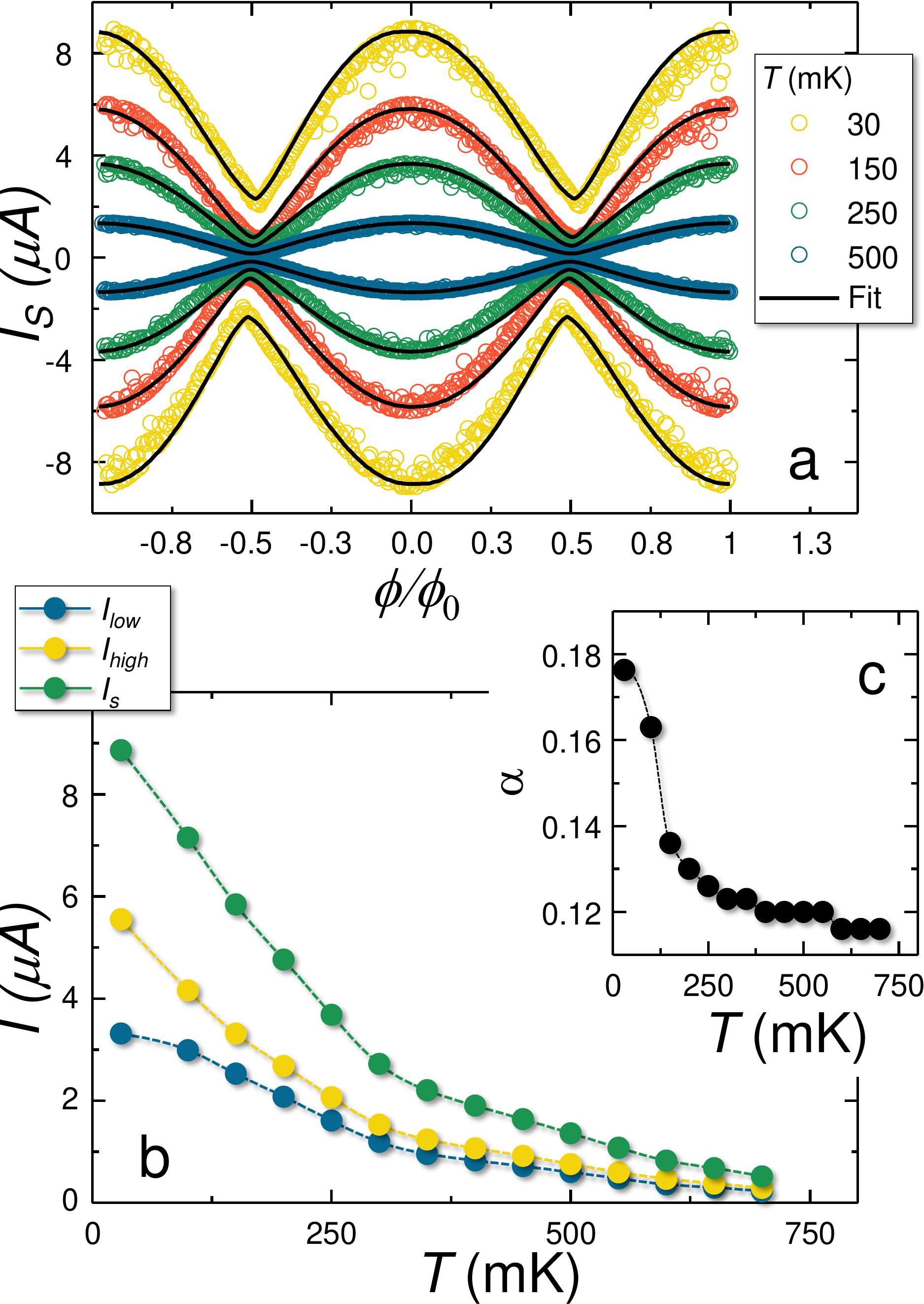}
    \caption{\textbf{Switching current \textit{vs.} flux characterization of the SNS SQUID.} 
    \textbf{a}: Switching current $I_S$ of the SNS SQUID as a function of the external magnetix flux $\phi$.  $\phi$ was applied through a superconducting electromagnet. $I_S(\phi)$ is shown for selected temperatures ranging between 30 mK and 500 mK. Superimposed on experimental data we show (black solid lines) the result of the fit obtained with the RSJ model. \textbf{b}: Plot of the maximum switching current of the SNS SQUID ($2I_0$)  versus temperature (green dots). In the same plot we also show the critical current of the two junctions extracted through the RSJ fit. 
    The value of the lowest of the two critical currents ($I_{low}$) is represented by blue dots, whereas the value of the highest of the two critical currents ($I_{high}$) is represented by yellow dots. \textbf{c}:  Asymmetry parameter ($\alpha$) as a function of  temperature. This value is extracted through the fitting procedure (see text).}
  \label{fig:fig2}
\end{figure}

Figure  \ref{fig:fig1}c  shows  the $IV$ characteristics of a representative SNS SQUID collected at  several  temperatures ranging  from  30  mK  to  650  mK.  The  curves  are  horizontally  offset for clarity.  For temperatures smaller than 750mK, the $IV$s exhibit a clear Josephson effect with a switching current $I_S$ of $\sim 7$ $\mu$A at 30 mK, and a normal-state resistance $R_N\sim 50 \Omega$.  Due to electron heating in the the normal state \cite{Courtois2008,Dubos2001}, the usual hysteretic  behavior  is  observed when  the $IV$ is measured  forward  and  backward with a retrapping current $I_R\sim2$ $\mu$A at 30 mK. 
A  plot of the switching and the retrapping current \textit{vs.} temperature ($T$) is  shown  in  Fig. \ref{fig:fig1}d. 
The difference between $I_S$ and $I_R$ decreases by increasing $T$, as routinely observed in similar systems \cite{Courtois2008,Dubos2001}, and vanishes at $T\sim350$ mK.

To study the $I_S(\phi)$ characteristics of  the SNS interferometers,  we  measured their $IV$s   as a function of the external magnetic field threading the SQUID loop. 
The device switching current was then extracted from the $IV$s to build the $I_s$ \textit{vs} $\phi$ curves. 
The $I_S(\phi)$ of the device is reported in Fig. \ref{fig:fig2}a for selected temperatures between 30 mK and 500 mK, where $\phi_0\simeq 2.067 \times 10^{-15}$ Wb is the magnetic flux quantum. 
For each temperature we plot both the positive ($I_{S_+}$) and negative  ($I_{S_-}$) switching current branches, defined accordingly to the scheme of Fig. \ref{fig:fig1}a. 
By defining the modulation amplitude $\Delta I_C=(I_{MAX}-I_{MIN})$ and the modulation average value  $<I>=(I_{MAX}+I_{MIN})/2$ (where $I_{MAX}$ and $I_{MIN}$ are the maximum and minimum value of $I_{S_+}$ respectively) a  modulation  visibility  $\Delta I_S  / <I> \sim 90 \% $, is observed at 30 mK. Such a value is on par with  the performance of state of the art SNS interferometers \cite{Angers2008,Ronzani2013}. 
Furthermore, it is about 9-fold higher than in gated monolithic Ti SQUIDs\cite{Paolucci2019a}. By increasing the temperature, both $<I>$ and $\Delta I_S$ decrease, due to
weakening of the proximity effect in the weak links, 
as routinely observed in these systems. 

The modulation visibility is mainly determined by the difference between the critical currents of the two junctions. The latter can be extracted by fitting the $I_S(\phi)$ data against the static zero-temperature resistively-shunted junction (RSJ) model\cite{Clarke2004}
\begin{equation}
    i=I_O [(1-\alpha) \sin(\delta_1)+(1+\alpha) \sin(\delta_2)],
\end{equation}
\begin{equation}
    2j=I_O [(1-\alpha) \sin(\delta_1)-(1+\alpha) \sin(\delta_2)],
\end{equation}
\begin{equation}
    \delta_2 - \delta_1=2\pi \phi/\phi_0 + \pi \beta j,
\label{quantization}
\end{equation}
where $\delta_{1,2}$ are the phase differences across the weak links, $i$ and $j$ are the supercurrent passing through and circulating in the SQUID, respectively. 
Within this formalism, defining $\alpha=\frac{|I_L-I_R|}{I_L+I_R}$, the asymmetry between the critical currents of the two junctions is accounted for. 
At fixed magnetic flux, $I_{S_+}$ and $I_{S_-}$ are defined as proportional respectively to the maximum and minimum values of $i$ over all the values of $\delta_1$ and $\delta_2$ satisfying Eqs. 1, 2, and 3, via the coefficient $I_0=\frac{I_L+I_R}{2}$, corresponding to one half of the maximum supercurrent of the SQUID as function of $\phi$. 
This model accounts also for the inductance $\mathcal{L}$ of the SQUID, through the screening coefficient $\beta=2\mathcal{L} I_0 /\phi_0$. 
Although the RSJ model was conceived for tunnel-like  Josephson junctions, it retains its validity also for SNS weak-links that, like ours, fall in the \textit{long} junction limit. 
A detailed description of the fit procedure is reported in the \textit{Methods} section. 
The fit curves are shown on top of experimental data in Fig. \ref{fig:fig2}a (solid black lines). 
The good agreement between the RSJ model and experimental data is quantitatively confirmed by the coefficient of determination $R^2$ of the fits, which ranges from 0.996 ( at 500 mK) to 0.97 (at 30 mK). 
The value for $2I_0$ determined through the fitting procedure is plotted against the temperature in Fig. \ref{fig:fig2}b. 
Furthermore, we extracted the $\alpha$ parameter, which is reported in Fig. \ref{fig:fig2}c. $\alpha$ reaches the maximum  value of $\sim0.2$ at 30 mK, and  decays when the temperature is increased.  
From $\alpha$ it is also possible to deduce the value of the critical currents of the two weak-links, that are $I_{high}\sim6$ $\mu$A and $I_{low}\sim3$ $\mu$A for the junction with the higher and the lower critical supercurrent, respectively. A plot of $I_{high}$ and $I_{low}$  as a function of the temperature is reported in Fig. \ref{fig:fig2}b. The value for $\beta$ derived from the fit is around 0.01 for every temperature, thereby confirming the negligible inductance contribution provided by the Al loop.

\begin{figure*}[t]
  \includegraphics[width=\textwidth]{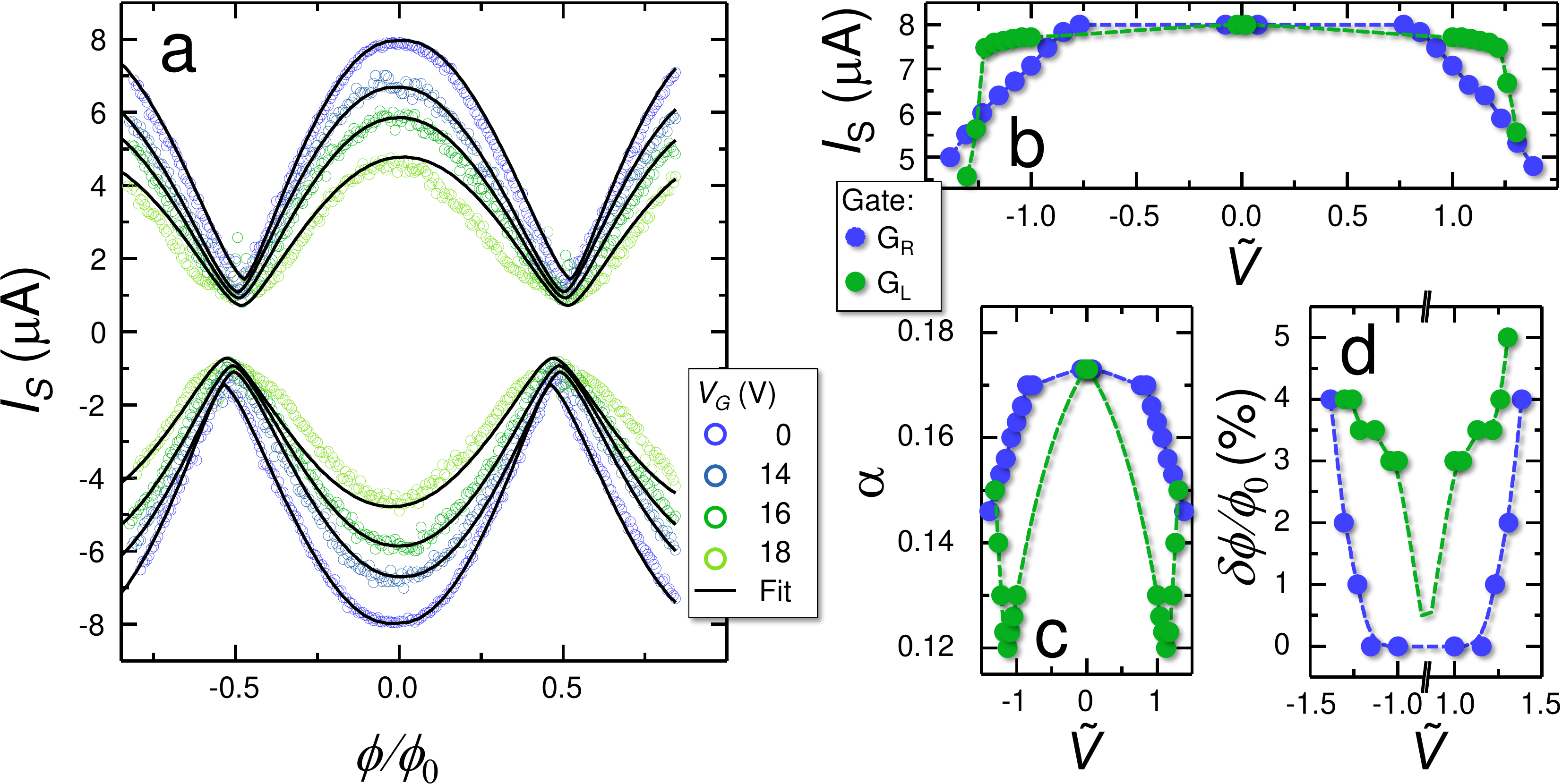}
  \caption{\textbf{Electrostatic control of the current-flux relation in a gated SNS SQUID.}  \textbf{a}: Positive ($I_{S_+}$) and negative ($I_{S_-}$) switching currents \textit{vs.} external flux of the SNS SQUID when a gate voltage is applied to one of the two junctions ($G_R$). 
  Curves are shown for gate voltage $V_R$ between 0  (unperturbed case) and 18 V. The minima of the interference pattern are almost locked within the explored voltage range. The maxima are dumped by a factor reaching $\sim 0.5$ at 18 V (green dots). Superimposed to experimental data we show (black lines) the result of the RSJ fit. At 18 V a significant deviation from the conventional single-tone behavior in the current-flux relation is observed. The same qualitative behaviour was observed when the left gate was polarized, and for negative values of the gate voltage. \textbf{b}: Maximum of the switching current of the SQUID \textit{vs}  normalized gate voltage $\tilde V$ applied either to the left (green dots) or right (blue dots) gate electrode. The normalization factors are 14 V and 46 V for the right and left gate, respectively, and correspond to a $\% 10$ suppression of the maximum switching current. \textbf{c}: Asymmetry parameter $\alpha$ as a function of the normalized gate voltage $\tilde V$ applied to the left (green) or right (blue) gate. 
  \textbf{d}: Plot of the additional phase shift ($\delta \phi$) introduced in the flux quantization relation (see Eq. \ref{quantization}) versus normalized voltage applied to the left (green) or right (blue) gate electrode.}
    \label{fig:fig3}
\end{figure*}

To investigate the impact of the gate bias on the SNS SQUID current-flux relation we measured $I_S(\phi)$ when several values of gate voltage were independently applied to either the left and right gate electrode. Figure \ref{fig:fig3}a shows the modulation patterns of $I_{S+}$ and $I_{S-}$  for different positive values of gate voltage $V_R$ applied to $G_R$ measured at 30 mK. $G_L$ was left grounded. 
It is worth to discuss the several interesting gate-dependent features emerging from the data. $I_{MAX}$ is constant up to about $12$ V. 
Above this threshold it is suppressed by further increasing $V_R$, and exhibits the same reduction for positive and negative gate voltages as well as for positive and negative current bias. 
The same qualitative behavior was observed by polarizing the left gate electrode, which due to a larger gate-junction distance was effective at higher voltages. 
The SQUID switching current as a function of gate voltage applied alternatively to $G_L$ or $G_R$ is shown in Fig. \ref{fig:fig3}b against $\tilde V$, i.e., the voltage normalized to the values at which the switching current was suppressed by $\%10$. This equals  14 V and 46 V for $G_R$ and $G_L$, respectively. 

In stark contrast to the conventional $T$-dependent case, in which both  minima and  maxima of the modulation pattern converge to 0 by enhancing the temperature, in the gate-dependent case the amplitude of the minima of $I_{S+}$ (and the maxima of $I_{S-}$)  are apparently almost locked in the explored voltage range.
We start our discussion on such an unconventional phenomenology by recalling that, following from Eq. 1,
\begin{equation}
I_{S+}(\phi)=I_0\sqrt{(1-\alpha)^2+(1+\alpha)^2+2(1-\alpha^2) \cos(\frac{2\pi \phi}{\phi_0})}.
\end{equation}
This expression, which holds when $\beta$ is negligible, allows to derive the $I_S(\phi)$ extremal values $I_{MAX}=2 I_0=I_L+I_R$ and $I_{MIN}=2 \alpha I_0=|I_L-I_R|$. These relations imply that it is not possible to affect $I_{MAX}$ (which in our data is suppressed by a $\sim50\%$ factor) keeping $I_{MIN}$ constant unless $I_L(V_R=0)-I_L(V_R)=I_R(V_R=0)-I_R(V_R)$ [and $I_R(V_L=0)-I_R(V_L)=I_L(V_L=0)-I_L(V_L)$] for each value of $V_R$ (and $V_L$). 
This condition is not only extremely unlikely to be satisfied, but it seems also incompatible with the typical length scale of the gating effect in metallic superconductors. Indeed, it was shown \cite{DeSimoni2018,Ritter2021} that the critical current suppression due the application of a gate voltage exponentially decays with the distance from the gate itself. 
In other words,   gating is a \textit{local} effect, which, acting on just one of the weak links, yet is able to affect \textit{non-locally} the response of the whole SNS SQUID.

In order to further elaborate on the above question, we believe  interesting to discuss  the results of the RSJ fitting of the $I(\phi)$s obtained at different gate voltage  values (see black lines in Fig. \ref{fig:fig3}a). 
The fit was performed by exploiting the same technique of the temperature-dependent case, but now including an additional phase shift ($\delta \phi$) in  Eq. \ref{quantization} such that $\delta_2 - \delta_1=2\pi \phi/\phi_0 + \pi \beta j + \delta \phi$. The introduction of the latter parameter was necessary to successfully fit the $I_{S\pm}(\phi)$ obtained for  $V_R > 15$V (and for $V_L > 50 V$). $\delta \phi$ is plotted as a function of the gate voltage applied to either $G_L$ or $G_R$ in Fig. \ref{fig:fig3}d, while $\alpha(\tilde{V})$ is plotted  in Fig. \ref{fig:fig3}c. For $|\tilde V| \lesssim 1$, the agreement between fit and data is optimal, with $R^2$ ranging between 0.98 and 0.99. Above this threshold, however, the ability of the RSJ model to represent the current-flux characteristics  progressively weakens: in particular, at $|\tilde V|=1.3$ (equivalent to $V_R = \pm 18$ V for the data represented in Fig \ref{fig:fig3}a), the deviation from the sinusoidal behavior is particularly evident. 
This behavior may be ascribed to a gate-induced modification of the Josephson current-phase relation of the weak-link, which is driven out from the conventional monochromatic regime, and turns out to be \textit{ coloured} with additional higher-harmonic terms. In the same voltage range, $\delta \phi$ increases reaching the maximum value of $\sim 0.04 \phi_0$ at $|\tilde V| \sim 1.5$.
It is also  worthwhile to  discuss the evolution of $\alpha(\tilde V)$, which monotonically decreases for increasing values of $|V_R|$ (blue dots in Fig. \ref{fig:fig3}c), while it shows minima for $\tilde V \sim \pm 1$ when $G_L$ is polarized (green dots in Fig. \ref{fig:fig3}c). In the framework of the RSJ model, such a behavior is equivalent to a simultaneous modification of the critical currents of both the weak-links, joint to the addition of a phase shift term. We stress that this characteristics is not compatible with a \textit{local} action of the gate voltage on the amplitude of the current-phase relation of the gated weak-link. 
Indeed, if this were the case, on the one hand, by gating the weak-link with the highest critical current, $\alpha$ should vanish (when $I_{high}=I_{low}$) and then increase up to 1. On the other hand, $\alpha$ is expected to monotonically converge to 1 when $I_{low}$ is suppressed, due to the enhancement of asymmetry between the two junctions. 
For this reason we hypothesize a voltage-driven modification of the phase-drop in the gated weak-link. This then affects  also the other weak-link, and  therefore the whole SQUID current-flux relation, through the flux quantization relation.

\section{Phase frustration through $\pi$-domain activation}
\begin{figure*}[t]
  \includegraphics[width=\textwidth]{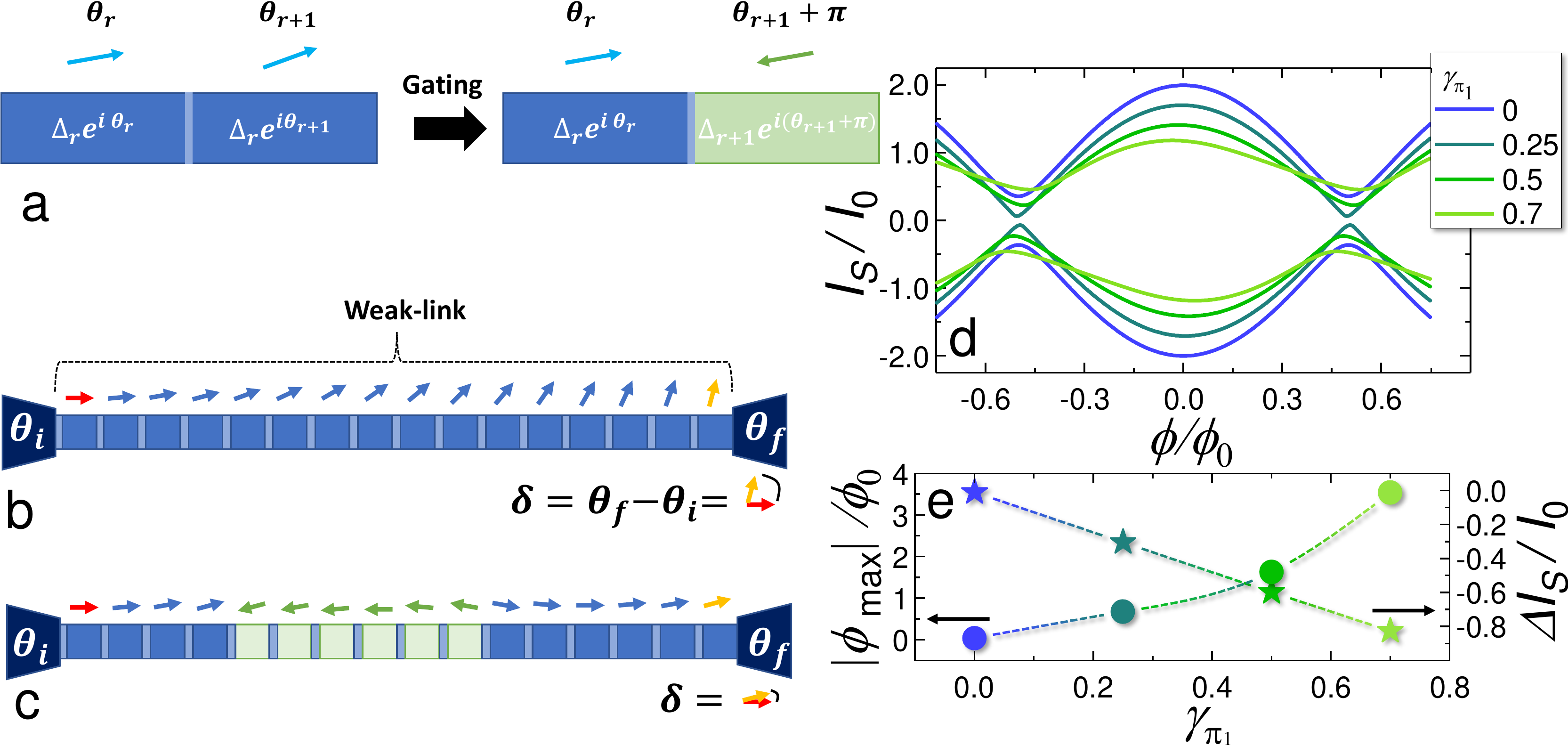}
    \caption{
    \textbf{Gate-driven phase frustration through $\pi$-rotation.}
    \textbf{a}: Pictorial representation of the $\pi$ rotation mechanism induced by the gate voltage. The superconductor is represented through a one-dimensional chain of domains (blue blocks); each of them can be described by a generic complex order parameter $\Delta_r e^{i \theta_r}$, where $r$ is a domain index, and $\theta_r$ is the superconducting phase in each domain. 
    Under the action of the gate voltage we assume the phase of some of the domains to be rotated by $\pi$  (green block).  \textbf{b}: When a supercurrent is injected through the weak-link, the phase drop ($\delta$) built across the latter results from the accumulation of phasor rotations acquired at each domain.  
    \textbf{c}: When a gate-voltage is applied, a fraction of the domains proportional to the intensity of the electric field acquire a phase-rotation of $\pi$ with respect to 0 V case. The resulting phase drop over the weak-link turns out to be frustrated due the counter-rotation acquired by the phasor in the $\pi$-domains. 
    \textbf{d}: $I_S(\phi)$ calculated through     Eq. \ref{frustration}  with $N=2$, $I_1=1.18$,  and $I_2=0.82$. The latter values correspond to an asymmetry parameter $\alpha=0.18$, \textit{i. e.,} equivalent  to that of our SNS SQUID. 
    The amplitude of $0$-phase harmonics $\gamma_{0_1}$ and $\gamma_{0_2}$ were respectively set to 1 and 0, in order to recover the conventional sinusoidal monochromatic behaviour when no gate voltage is applied. We shows curves obtained for $\gamma_{\pi_2}=0.05 \gamma_{\pi_1}$ and  for selected values of $\gamma_{\pi_1}$ ranging between 0  and 0.7. The former case corresponds to an unpolarized gate voltage. \textbf{e}: Phase shift of the maxima of $I(\phi)$, $\phi_{MAX}$, as a function of  $\gamma_{\pi_1}$ (left axis). 
    The reported values are consistent with the result of the RSJ model for the experimental data for the parameter $\delta \phi$. 
    We also show the $\gamma_{\pi_1}$-dependence of the normalized variation of the maximum value of the interference pattern $\Delta I_S/I_0$ (right axis).}
  \label{fig:fig4}
\end{figure*}

We now discuss a possible phenomenological  model based on the assumption that the gate voltage is able to affect only the phase of each superconducting domain composing the weak-link, and rotating it by a factor of $\pi$ (see Fig. \ref{fig:fig4}a). This hypothesis assumes the existence of a fully coherent mechanism that  is able  to account for all the main features observed in gate-controlled $I_S(\phi)$s. Due to the polycristalline nature of the copper wire forming our weak-links, we describe each domain through an order parameter $\Delta_r e^{i\theta_r}$, where $\Delta_r$ and $\theta_r$ are the amplitude of  the gap and phase of the $r^{th}$ domain, respectively \cite{Tinkham2004}. In this framework, when a supercurrent is injected through the weak-link, the phase drop $\delta$ built across the latter results from the accumulation of the phasor rotations acquired at each domain (see Fig. \ref{fig:fig4}b). 
In this condition, the current-phase relation of the weak-link can be described by the conventional Josephson equation $I=I_1 sin\delta$, where $I$ is the biasing current. 
When a gate voltage is applied, a fraction of the domains proportional to its intensity acquires a phase rotation of $\pi$ (see Fig. \ref{fig:fig4}c) with respect to the unperturbed value. The phase drop over the weak-link in this configuration is, therefore, overall \textit{frustrated} due the counter-rotation acquired by the phasor in the $\pi$-domain (green blocks in Fig. \ref{fig:fig4}b). 
This physical intuition finds a mathematical representation by modifying the weak-link current phase relation as follows:
\begin{align}
\begin{split}
    I & = I_1 \left [ \sum_{n=1}^N \gamma_{0_n} \sin (n \delta )+ \gamma_{\pi_n} \sin (n \delta +  \pi ) \right ]=\\
    & = I_1 \sum_{n=1}^N (\gamma_{0_n}-\gamma_{\pi_n}) \sin (n \delta ),
\end{split}
\end{align}
where we recover the most general functional form \cite{Heikkila2002} by including an arbitrary number N of 0-phased and $\pi$-phased harmonics with weight $\gamma_{0_n}$ and $\gamma_{\pi_n}$, respectively, determined by the contribution of the $\pi$ domains to the resulting phase drop. Following from this assumption, the RSJ current-flux relation of the SQUID modifies into
\begin{equation}
    i =I_1  \sum_{n=1}^N (\gamma_{0_n}-\gamma_{\pi_n}) sin (n \delta )+I_2 sin\left(\delta+\frac{2\pi\phi}{\phi_0}\right),
\label{frustration}
\end{equation}
where $I_1$ and $I_2$ account for the amplitude of the critical current of gated and non-gated weak-link, respectively. 
Figure \ref{fig:fig4}d shows the $I_S(\phi/\phi_0)$ calculated through this model with just two harmonics ($N=2$), and with $I_1=1.18$  and $I_2=0.82$. The latter values correspond to an asymmetry parameter $\alpha=0.18$, \textit{i. e.,} compatible  to that of our SNS SQUIDs. The amplitude of $0$-phase harmonics $\gamma_{0_1}$ and $\gamma_{0_2}$ were set respectively to 1 and 0 in order to recover the conventional sinusoidal \textit{monochromatic} behaviour when no gate voltage is applied. We show curves obtained for $\gamma_{\pi_2}=0.05 \gamma_{\pi_1}$ and  $\gamma_{\pi_1}$ ranging between 0 (blue curve in Fig. \ref{fig:fig4}d),  and 0.7 (light-green curve). 
The former case corresponds to a vanishing gate voltage. 
By increasing $\gamma_{\pi_1}$, we mimic the action of the gate voltage, which amplifies the weight of the $\pi$ terms for both the harmonics, thereby resulting in a suppression of the maxima of the current-flux relation (star plot in Fig. \ref{fig:fig4}d). The latter reaches a value of $\sim 50 \%$ for $\gamma_{\pi_1}=0.7$. Besides, $I(\phi)$ minima undergo a non-monotonic and much more limited variation. 
We wish to emphasize that, by introducing just one additional harmonic, we obtained  a significant deviation from the sinusoidal behavior, which resembles that of the experimental data. 
Furthermore, the shift of the maxima of $I_S(\phi)$ (dots in Fig. \ref{fig:fig4}d) are consistent with the result of the RSJ fit procedure for the parameter $\delta \phi$, reaching a value of $\sim 4\%$ (see  Fig. \ref{fig:fig3}d for a comparison).

\section{Effect of gating in the dissipative regime}
Among available magnetic field sensors, SQUIDs are the devices of choice for those applications requiring ultra-high sensitivity at the nanoscale. SQUIDs have progressively become an essential tool for probing several systems, such as magnetic molecules and nanoparticles, single electrons, and cold atom clouds. Beyond the detection of magnetic moments (down to the single spin resolution),  SQUIDs play in a front row role in a vast  field of applications ranging from microbolometry\cite{Giazotto2008a} and spintronics to drug delivery and cancer treatment. In this last section we discuss the performance of our SNS SQUID in view of its possible exploitation as a gate-tuned magnetic flux sensors operating in the dissipative regime. 
The latter is conventionally obtained by current biasing the interferometer above its critical current.  Variations of the magnetic field threading the loop translate into variations of the voltage drop ($V$) developed across the Josephson junctions.
\begin{figure*}
  \includegraphics[width=\textwidth]{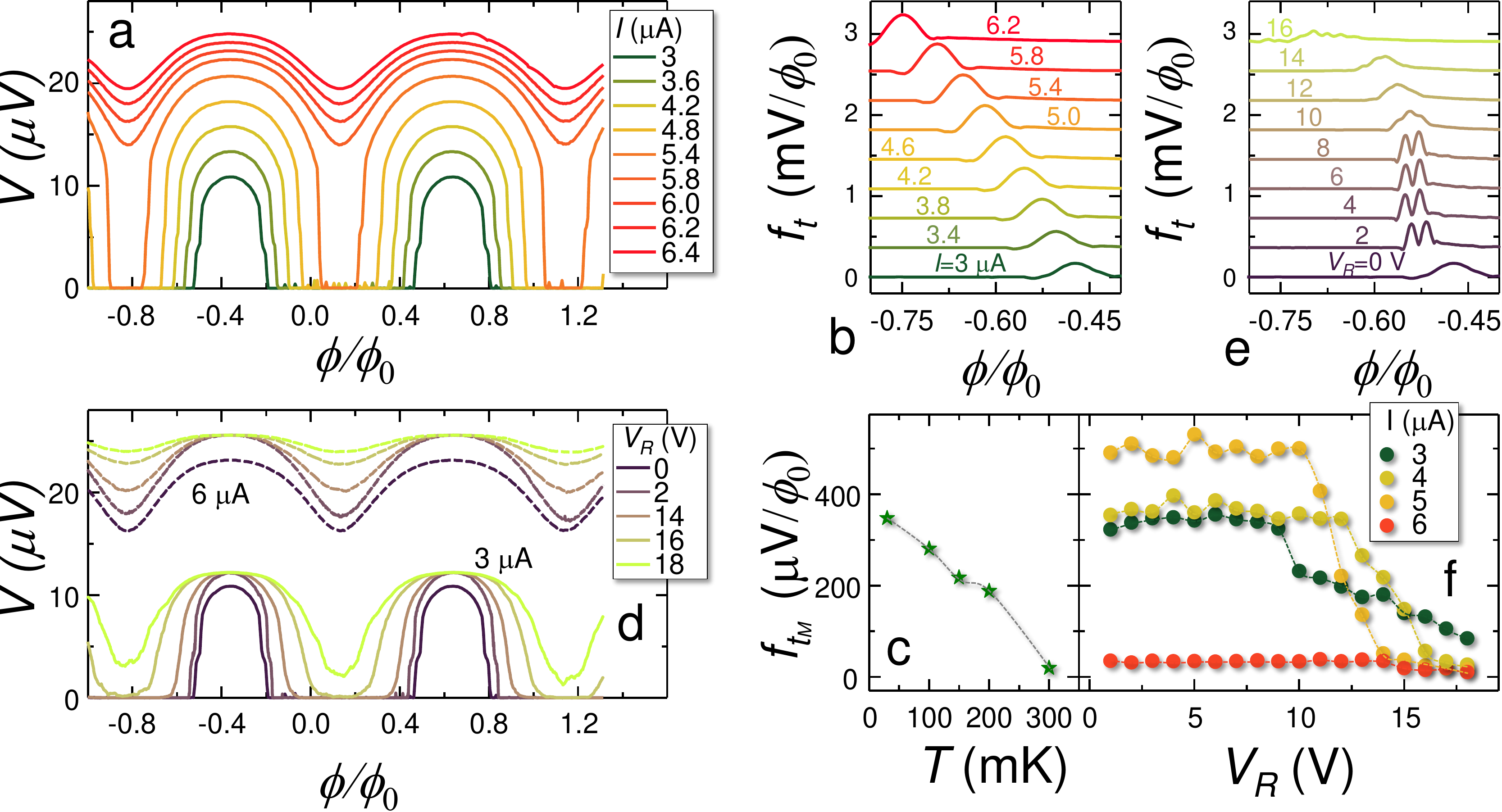}
    \caption{
    \textbf{Effect of gating in the dissipative regime.}
    \textbf{a} $V(\phi)$ characteristics at 30 mK for selected bias current values between 3 $\mu$A and 6.4 $\mu$A. The measurements were performed with a standard 4-wire lock-in technique by biasing the device via a 17 Hz sinusoidal current signal. Below $I\simeq 6$ $\mu$A, the curves exhibit a zero voltage-drop for magnetic fluxes such that $I<I_C(\phi)$. A finite $V$ value is, instead, measured when the device switches into the dissipative regime due to bias current being higher than the flux-dependent critical current.
    \textbf{b} Transfer function $f_t$ vs $\phi$ for selected amplitudes of the biasing current. $f_t$ was calculated through numerical differentiation of the $V(\phi)$ characteristics measured at 30 mK.
    \textbf{c}: Maximum value of $f_t$ ($f_{t_M}$) \textit{vs} $T$. 
    \textbf{d}: $V(\phi)$ curves obtained for $I=3$ $\mu$A (solid lines) and $I=6$ $\mu$A (dashed lines) and for $V_R$ ranging between 0 and 18 V.
    \textbf{e} Transfer function $f_t$ vs $\phi$ for selected values of $V_R$ at $30$ mK.
    \textbf{f}: $f_{t_M}$ \textit{vs} $V_R$ for selected values of the biasing current at 30 mK.}
  \label{fig:fig5}
\end{figure*}
Figure \ref{fig:fig5}a shows the $V(\phi)$ curves measured at 30 mK on a representative device by 4-wire lock-in technique for selected amplitudes of the 17 Hz sinusoidal current-bias signal $I$. 
Below  $I\sim 6\mu$A, the curves exhibit  a zero voltage-drop for magnetic fluxes such that $I<I_S(\phi)$. A finite $V$ value is, instead, measured when the interferometer switches into the dissipative regime, for the biasing current being higher than the flux-dependent switching current. This results in a strongly nonlinear behaviour at the switching points, corresponding to a high value for the flux-to-voltage transfer function, $f_t=|\partial V / \partial \phi|$. The latter characteristic, nonetheless, cannot be easily exploited for highly-sensitive operation due to the stochastic nature of the switching, which results in an unstable working point and in a vanishing dynamic range \cite{Ronzani2013}. The transfer function, calculated through numerical differentiation of the $V(\phi)$ curves, is shown in Fig. \ref{fig:fig5}b for selected bias current values. The current provides an useful knob to select the flux values at which the interferometer responsivity is maximized. 
The maximum value of $f_t$ ($f_{t_M}$) is plotted \textit{vs} $T$ in Fig. \ref{fig:fig5}c. 
$f_{t_M}$ decreases with the temperature almost linearly from the value of 400 $\mu$V/$\phi_0$ obtained at 30 mK, and vanishes around $300$ mK. Such a performance is on par with that of interferometers of similar  typology \cite{Ronzani2013}.

The impact of the gate voltage on the $V(\phi)$ was explored by repeating the acquisition of such characteristics at 30 mK as function of both the current bias and the voltage applied to either $G_L$ or $G_R$. Figure \ref{fig:fig5}d shows the $V(\phi)$ curves obtained for $I=3$ $\mu$A (solid lines) and $I=6$ $\mu$A (dashed lines) for $V_R$ ranging between 0 and 18 V. The first family of curves (3 $\mu$A) corresponds to a position in the parameters space where, at null gate voltage, the interferometer is not fully operated in the dissipative regime. 
The  0-voltage-drop flux interval was observed to shrink by increasing the intensity of the gate voltage, until it completely disappears due to the gate-driven suppression of the critical current of the SQUID. 
At $V_R=18$ V the device operates in a fully-dissipative regime. 
The second family of curves (6 $\mu$A) falls entirely in the dissipative regime. We note that the result of the action of the gate is rather different from
 the behavior obtained by increasing the biasing current. Indeed,  in the latter case both the minimum and the maximum of the modulation pattern increases by increasing the bias current. In the gate-driven regime, instead, the maximum of the modulation turns out to be  locked, whereas the minimum can be controlled through the gate. These characteristics can be exploited to adapt to specific tasks the transfer function of the interferometer at the switching points through an additional knob, the gate voltage. 
 By shrinking  the width of the non-dissipative region through the gate action, \textit{e. g.}, it is possible to magnify the flux dynamic-range at the switching point without reducing the overall voltage-drop swing and the resulting device sensitivity. The plot of $f_t$ {vs.} $\phi$ at $I=3$ $\mu$A and $T=30$ mK is shown in Fig. \ref{fig:fig5}e for several values of $V_R$. We note that $f_{t_M}$  remains almost constant in a wide gate-voltage range, as shown in Fig. \ref{fig:fig5}f for selected bias current values.

\section{Conclusions}
The physics of electrostatic gating on metallic superconductors is, to date, one of the latest unanswered questions in condensed matter physics. Despite a few theoretical interpretations have been proposed, a model able to account for the totality of the phenomenology observed so far, and to provide quantitative prediction has not been developed yet. 
Our experiments on gated all-metallic SNS SQUIDs show that the microscopic mechanism at the origin of the    critical current suppression of gated weak-links is apparently phase coherent and produces a softening of the  phase rigidity of the Josephson junctions. This latter observation provides a  valuable reason to exclude any thermal-assimilated origin of gate-driven effects, on the one side. 
On the other side, we claim that, among the models aiming at the description of electrostatic gating in metallic superconductors, those in which it will be possible to take into account phase coherent effects should be preferred. 
Here, we interpreted our data through a phenomenological model based on the sole assumption that the gate induces a phase rotation of $\pi$ in the superconducting domains of the weak-link subjected to the action of the electric field. 
Although  rather simplified, our model successfully captures the main features observed in gated all-metallic SNS SQUIDs, such as the suppression of the maximum switching current, the blocking of the minimum switching current, and the deviation from the monochromatic beahvior of the interferometer current-flux relation.
We conclude by emphasizing the practical relevance of gated all-metallic SNS SQUIDs for magnetometry at the nanoscale. Indeed, the gate voltage provides an additional control on the transfer function of the interferometer, which can be exploited to tailor the response of the device on specific needs such as, for instance, the amplification of the flux dynamic range around the switching points for applications requiring  higher sensitivity.


\section*{Methods}
\subsection{Device nanofabrication}
The SNS-SQUIDs were fabricated by a single-step electron-beam lithography (EBL) and two-angle shadow-mask metal deposition through a suspended resist-mask onto an intrinsic Si(111) wafer covered with 300 nm of thermal SiO$_2$. The metal-to-metal clean interfaces were realized at room temperature in an ultra-high vacuum (UHV) chamber (base pressure $\sim5\times 10^{-11}$ Torr) of an electron-beam evaporator equipped with a tiltable sample holder. 
A 5-nm-thick Ti adhesion-film was deposited at an angle of $0^\circ$. Subsequently, 25 nm of Cu were evaporated to realize the SQUID nanowires and gates. Finally, the sample holder was tilted at $13^\circ$ for the deposition of a 100-nm-thick layer of Al to realize the superconducting loop.

\subsection{Cryogenic electrical characterization}
The  electrical  characterization  of  our  devices  was  performed  by  four-wire technique  in a  filtered cryogen-free $^3$He-$^4$He dilution  fridge equipped with a superconducting electromagnet, used to apply the external magnetic flux. Current-voltage ($IV$) measurement were carried out by setting a low-noise current bias and measuring the voltage drop across the weak-links with a room temperature pre-amplifier. Switching current average values were calculated over the switching points extracted from 15 repetitions of the same $IV$. The voltage-flux characterization was performed through a standard lock-in technique: the sinusoidal reference signal of the lock-in was used to current-bias the device. The in-phase output voltage signal was pre-amplified at room temperature. The gate voltage was applied through a room-temperature low-noise voltage source. The devices were also characterized in terms of gate-weak-link leakage current, which was found to be always lower than 1 pA.

\subsection{RSJ fit of experimental data}
The fitting procedure was based on Eqs. (1)-(3) together with the maximum condition $i_s^+=\max\limits_{\delta_1,\delta_2}(i), \quad i_s^-=\min\limits_{\delta_1,\delta_2}(i)$. Substituting Eqs. (2),(3) in Eq. (1) we obtain a function for the current through the loop, depending on the flux $\phi$, with $\alpha$, $\beta$, $I_0$ and $\delta_{1,2}$ as parameters.\\
The code used for the fit 
 minimizes the distance of the function from the experimental points.

\section*{Acknowledgement}
The authors  acknowledge the European Research Council under Grant Agreement No. 899315-TERASEC, and  the  EU’s  Horizon 2020 research and innovation program under Grant Agreement No. 800923 (SUPERTED) 
and No. 964398 (SUPERGATE)
for partial financial support.


\section*{Author Contributions}
N.L. fabricated the devices. S.B and G.D.S performed the experiment with input from F.G.. S.B. analyzed the data with input from G.D.S and F.G. G.D.S implemented the numerical model with inputs from M.T.M, M.C., and F.G. G.D.S. wrote the manuscript with input from all the authors. F.G. conceived the experiment. All of the authors discussed the results and their implications equally.

\providecommand{\latin}[1]{#1}
\makeatletter
\providecommand{\doi}
  {\begingroup\let\do\@makeother\dospecials
  \catcode`\{=1 \catcode`\}=2 \doi@aux}
\providecommand{\doi@aux}[1]{\endgroup\texttt{#1}}
\makeatother
\providecommand*\mcitethebibliography{\thebibliography}
\csname @ifundefined\endcsname{endmcitethebibliography}
  {\let\endmcitethebibliography\endthebibliography}{}

\end{document}